\documentclass[conference, 10pt]{IEEEtran}
\IEEEoverridecommandlockouts

\usepackage{cite}
\usepackage{amsmath,amssymb,amsfonts,amsthm}
\usepackage{graphicx}
\usepackage{textcomp}
\usepackage[dvipsnames]{xcolor}
\usepackage{braket}
\usepackage{hyperref}
\usepackage{subcaption}

\usepackage[linesnumbered,ruled,noend]{algorithm2e}

\makeatother

\newtheorem*{remark}{Remark}

\newcommand{\eqdef}{\stackrel{\triangle}{=}}

\def\BibTeX{{\rm B\kern-.05em{\sc i\kern-.025em b}\kern-.08em
    T\kern-.1667em\lower.7ex\hbox{E}\kern-.125emX}
}

\newcommand{\mpfont}{\scriptsize}

\ifx\noeditingmarks\undefined
    \newcommand{\MPworker}[2]{{\color{#1}\vrule\vrule}{\marginpar{\color{#1}\mpfont #2}}}
\else
    \newcommand{\MPworker}[2]{}
\fi

\usepackage[scr=boondox,  
            cal=esstix]   
           {mathalpha}

\begin{document}

\title{Simulation of Entanglement-Enabled Connectivity in QLANs using SeQUeNCe}

\author{
\IEEEauthorblockN{Francesco Mazza$^{*}$, Caitao Zhan$^{\dagger}$, Joaquin Chung$^{\dagger}$, Rajkumar Kettimuthu$^{\dagger}$, Marcello Caleffi$^{*}$,\\ Angela Sara Cacciapuoti$^{*}$}

\IEEEauthorblockN{$^{*}$University of Naples Federico II (Italy), $^{\dagger}$Argonne National Lab (USA)}

\thanks{*F. Mazza, M. Caleffi and A. S. Cacciapuoti are with the \href{www.quantuminternet.it}{www.QuantumInternet.it} research group, \textit{FLY: Future Communications Laboratory}, University of Naples Federico II, Naples, 80125 Italy. 
}
\thanks{*A. S. Cacciapuoti and F. Mazza acknowledge PNRR MUR NQSTI-PE00000023, M. Caleffi acknowledges PNRR MUR project RESTART-PE00000001.}
\thanks{$\dagger$ This material is based upon work supported by the U.S. Department of Energy, Office Science, Advanced Scientific Computing Research (ASCR) program under contract number DE-AC02-06CH11357 as part of the InterQnet quantum networking project.}
}

\maketitle

\begin{abstract}

    Quantum Local Area Networks (QLANs) represent a promising building block for larger scale quantum networks with the ambitious goal --in a long time horizon-- of realizing a Quantum Internet. Surprisingly, the physical topology of a QLAN can be enriched by a set of \textit{artificial links}, enabled by shared multipartite entangled states among the nodes of the network. This novel concept of artificial topology revolutionizes the possibilities of connectivity within the local network, enabling an on-demand manipulation of the artificial network topology. In this paper, we discuss the implementation of the QLAN model in SeQUeNCe, a discrete-event simulator of quantum networks. 
    Specifically, we provide an analysis of how network nodes interact, with an emphasis on the interplay between quantum operations and classical signaling within the network.
    Remarkably, through the modeling of a \textit{measurement protocol} and a \textit{correction protocol}, our QLAN model implementation enables the simulation of the manipulation process of a shared entangled quantum state, and the subsequent engineering of the entanglement-based connectivity. 
    Our simulations demonstrate how to obtain different virtual topologies with different manipulations of the shared resources and with all the possible measurement outcomes, with an arbitrary number of nodes within the network.
\end{abstract}


\section{Introduction}
\label{sec:1}

\begin{figure*}[t!]
    \centering
    \begin{subfigure}[b]{0.3\linewidth}
        \centering
        \includegraphics[width=\linewidth]{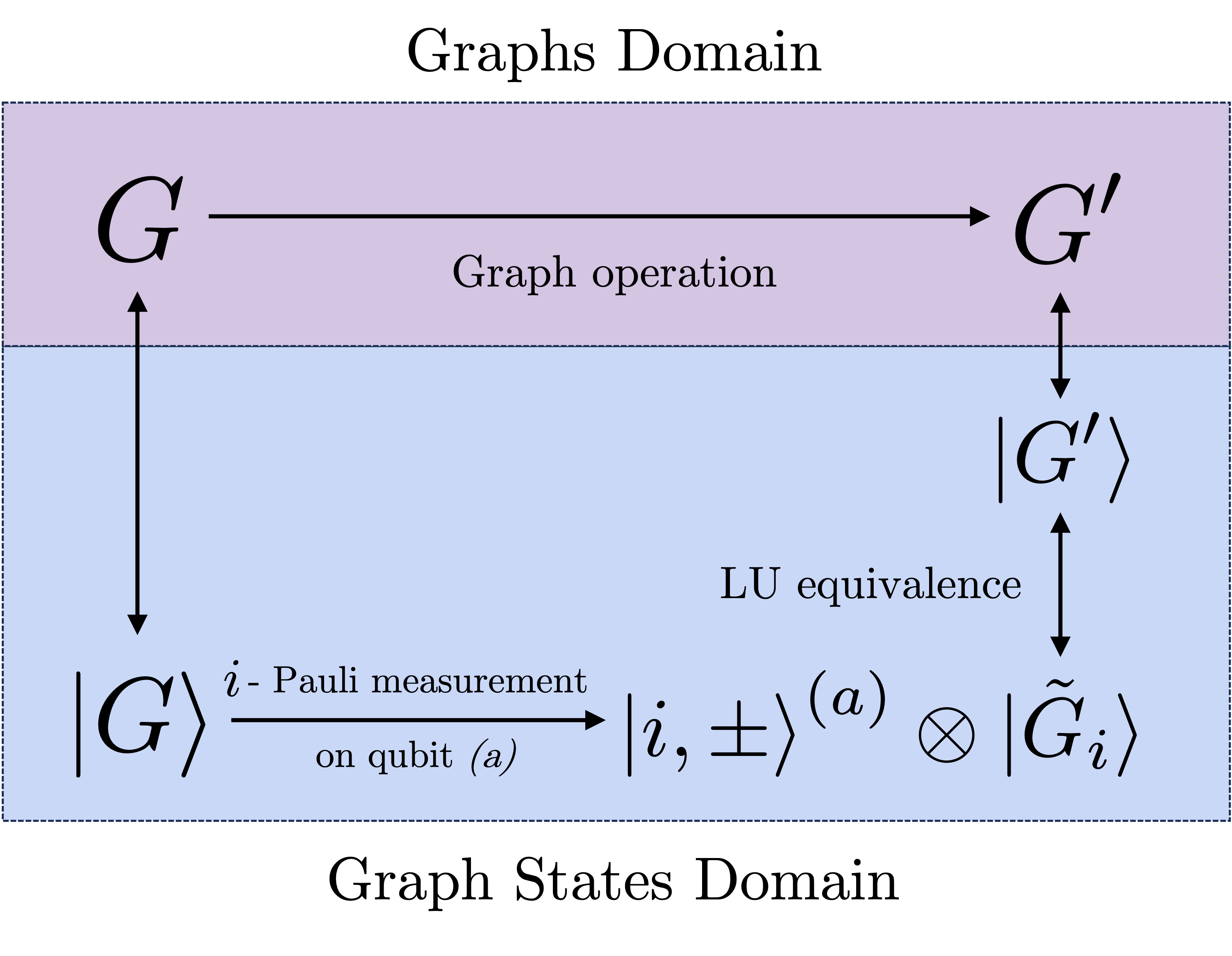}
        \caption{Schematic diagram of the correspondence between
        graph domain and graph state domain. Figure reproduced from \cite{CheIllCac-24}.}
        \label{fig:01a}
    \end{subfigure}
    \hfill
    \begin{subfigure}[b]{0.67\linewidth}
        \centering
        \begin{tabular}{c|c}
            \hline
            \hline
            \textbf{Resulting Graph state} & \textbf{Graph operations} \\
            \hline
            $P_{z,\pm}^{(a)} \ket{G}  = \ket{z,\pm}^{(a)} \otimes \underbrace{U_{z,\pm}^{(a)}\ket{G - a}}_{\ket{\tilde{G}_z}}$ & $\tilde{G}_z = G - a$ \\
            $P_{y,\pm}^{(a)} \ket{G}  = \ket{y,\pm}^{(a)} \otimes \underbrace{U_{y,\pm}^{(a)}\ket{\tau_a(G)-a}}_{\ket{\tilde{G}_y}}$ & $\tilde{G}_y = \tau_a(G) - a$ \\
            $P_{x,\pm}^{(a)} \ket{G}  = \ket{x,\pm}^{(a)} \otimes \underbrace{U_{x,\pm}^{(a)}\ket{\tau_{b_0}\left(\tau_{a}\left(\tau_{b_0}(G)\right) - a\right) }}_{\ket{\tilde{G}_x}}$ & $\tilde{G}_x = \tau_{b_0}\big( \tau_{a} (\tau_{b_0}(G)) - a\big)$ \\
            \hline
        \end{tabular}
        \caption{Table of the Pauli measurements and the corresponding graph operations, up to the application of the correction unitaries that depend on the outcomes \cite{HeiEisBri-04, HeiDurEis-06}.
        }
        \label{fig:01b}
    \end{subfigure}
    \caption{Schematic representation of the mapping between graph states and graph theory tools (a) under the application of Pauli measurements (b).}
    \label{fig:01}
    \hrulefill
\end{figure*}

As a cross-layer resource, entanglement has undoubtedly a pivotal role in the design of future quantum communication networks \cite{IllCalMan-22, MazCalCac-24, MazCalCac-24-QCNC, KozWehVan-22}. Specifically, the application of entanglement in its multipartite form is receiving considerable attention from the community \cite{DurVidCir-00, EisBriHan-01}, given its importance as an exceptional resource for revolutionizing the connectivity of quantum networks and enabling novel and surprising network functionalities \cite{HilGupZha-qsn, BriRau-01, CheLo-07, RamPirDur-21, IllCalVis-23, BarBirBom-23, ChuRamAni-24}.
Unlike classical networks, Quantum Local Area Networks (QLANs) can overcome physical network constraints through multipartite entanglement that allow flexible on-demand creation of \textit{virtual (or artificial) links}\footnote{virtual or artificial links between two QLAN nodes reflect the interaction pattern between the qubits belonging to the composite multipartite entangled state.} between nodes \cite{MazCalCac-24-QCNC, MazCalCac-24, CheIllCac-24}. 
Remarkably, artificial links constitute a sort of ``overlay entangled topology'' built upon the physical one, referred to as \textit{artificial topology}, that can differ significantly from the physical topology.
Specifically, by exploiting the unique properties of two-colorable graph states \cite{AshDurBri-05}, artificial topologies can be engineered to enhance network communication capabilities.

A theoretical framework for QLAN and the introduction of the necessary tools for network topology engineering were proposed in~\cite{MazCalCac-24, MazCalCac-24-QCNC}. Nevertheless, the practical realization of complex quantum communication scenarios is limited by hardware that is not yet sufficiently mature. Therefore, researchers resort to quantum network simulators to conduct empirical simulations to verify the theories and analytical results. In recent years, quantum network simulators~\cite{WuKolChu-21, netsquid_2021, quisp_2022} have played a crucial role in studying quantum network hardware~\cite{ZanKolChu22,ZhoLaiGan-23,SooBenHaj24}, protocols~\cite{Dahlberg_2019,KozDahWeh_2020, GhaZhaGup-22}, and applications~\cite{SunGupRam24,ZanKolGon24}. Motivated by the above, this paper addresses the implementation of the QLAN framework discussed in~\cite{MazCalCac-24} and~\cite{MazCalCac-24-QCNC} in a discrete-event simulator of quantum networks called SeQUeNCe~\cite{WuKolChu-21}. The contributions of this manuscript are the following:

\begin{enumerate}
    \item[i)] We extend SeQUeNCe by adding new classes of network node modules: orchestrator node and client node. These modules represent the core component of the QLAN system that enables the dynamic construction of artificial topologies by only performing Local Operations and Classical Communication (LOCC).
    
    \item[ii)] We design and implement a \textbf{\textit{measurement protocol}} and a \textbf{\textit{correction protocol}} that allow the simulation of the manipulation of shared entangled quantum states and the subsequent engineering of entanglement-based connectivity.
\end{enumerate}

More in detail, the two protocols provide an analysis of the network nodes' interaction with the interplay between quantum operations and classical signaling within the network. Remarkably, the conducted simulation studies demonstrates how various virtual topologies can be achieved through different manipulations of the shared resources, accounting for all possible measurement outcomes, with an arbitrary number of nodes in the network. The implemented software modules are open source and publicly available at \cite{GithubSequence}. To the best of our knowledge, this is the first paper to address the software implementation of the key functionality enabled by the QLAN, i.e., the engineering of the network topology through the manipulation of a shared multipartite resource.

\section{Preliminaries}
\label{sec:2}

\begin{figure*}[ht]
    \centering
    \includegraphics[width=0.68\linewidth]{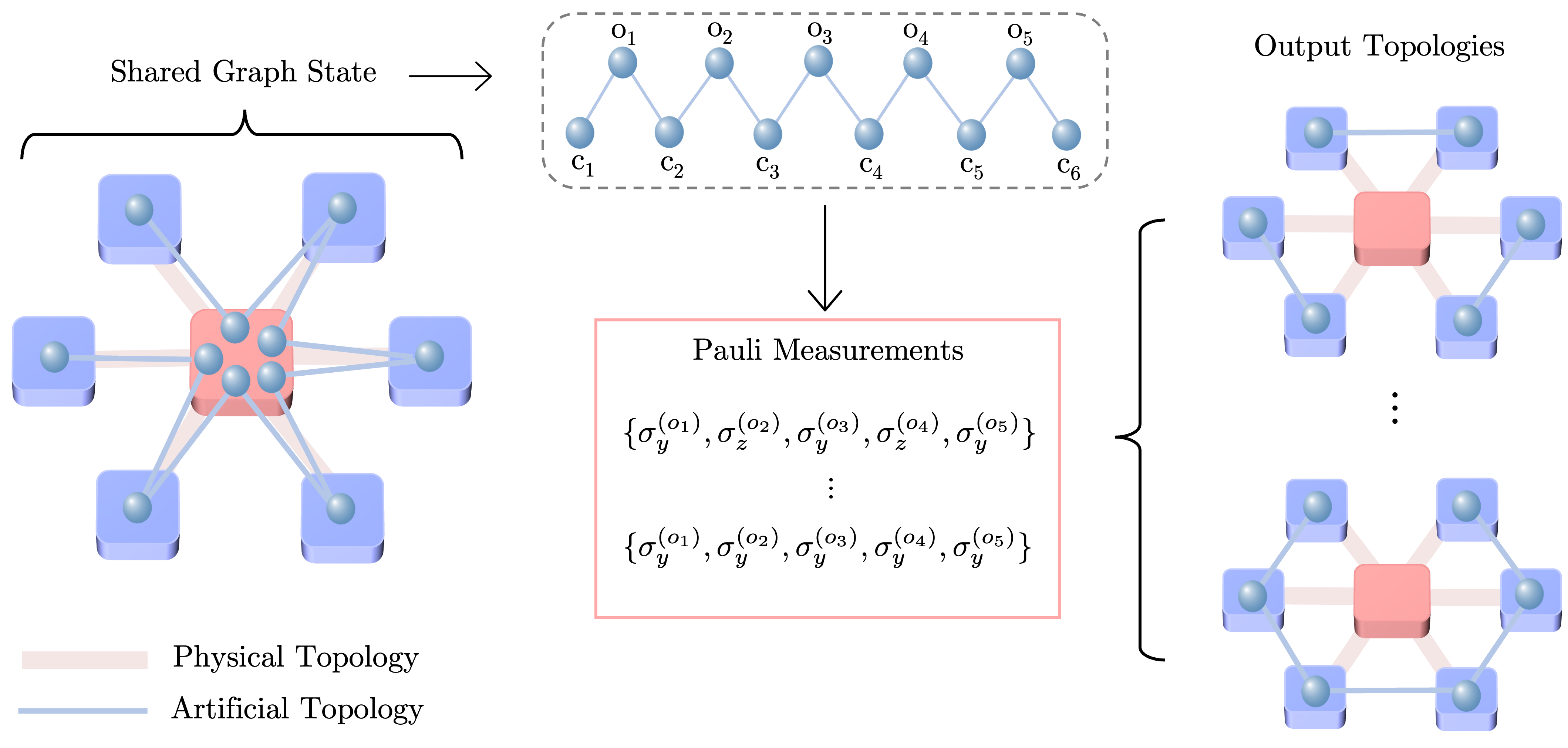}
    \caption{Some examples of the resulting graph states after the application of Pauli measurements and corrections.}
    \label{fig:02}
    \hrulefill
\end{figure*}

\subsection{Introduction to Graph states}
\label{sec:2.1}
Graph states gained relevant attention within the quantum communication community thanks to their unconventional properties \cite{HeiEisBri-04}. 
Specifically, graph states can be effectively described with graph theory tools and thus, any given graph state $\ket{G}$ corresponds to an associated graph $G$. 

Formally, a graph state $\ket{G}$ associated to the graph $G = (V,E)$ is defined as follows: 
\begin{equation}
    \label{eq:01}
	\ket{G} = \prod_{\{a,b\} \in E} \texttt{CZ}_{ab}\ket{+}^{\otimes n},
\end{equation}
with $\ket{+} = \tfrac{1}{\sqrt{2}}(\ket{0} + \ket{1})$, $n = |V|$ and $\texttt{CZ}_{ab}$ denoting the Controlled-Z gate applied to the qubits associated to the vertices $a$ and $b$. 
One of the most fascinating consequences of this formulation, also briefly presented in Fig.~\ref{fig:01a}, is that quantum operations -- such as Pauli measurements -- on the graph state can be associated with some simple operations on the associated graph G. 
In particular, the action of the Pauli projection operators on some qubits of the graph state $\ket{G}$ can be straightforwardly related to local complementation operations on the associated graph $G$ \cite{HeiEisBri-04,HeiDurEis-06}. 
Remarkably, as schematically represented in Fig.~\ref{fig:01b}, the application of a projection operator -- $P_x$, $P_y$, $P_z$ -- on a qubit of the graph state $\ket{G}$ corresponds, up to some local correction unitaries $U_{i,\pm}$, to a new graph state $\ket{\tilde G_i}$, on the remaining qubits, whose associated graph $\tilde G_i$ can be obtained through simple graph transformations on the associated graph $G$ of the initial state $\ket{G}$. The application of Pauli measurements on graph states is defined as follows:

\begin{enumerate}
    \item[I)] \textbf{Projective $\sigma_z$ Pauli measurements} \\
            The projective measurement via Pauli operator $\sigma_z$ of a qubit associated to the vertex $a$ of the graph $G$ of the initial graph state $\ket{G}$, yields to i) the deletion of the measured vertex on the associated graph $G$ and ii) the resulting graph state $\ket{\tilde G_z} \eqdef U_{z,\pm}^{(a)}\ket{G - a}$ on the unmeasured qubits. Specifically, the correction unitaries $U_{z,\pm}$ are defined according to the measurement outcome and take into account the neighborhood $N_a$ of the measured qubit $a$:
        \begin{align}
            \label{eq:02}
            U_{z,+}^{(a)} = \mathbb{I} \quad , \quad
            U_{z,-}^{(a)} = \bigotimes_{b \in N_a}\sigma_z ^{(b)}.   
        \end{align}

    \item[II)] \textbf{Projective $\sigma_y$ Pauli measurements} \\
        The projective measurement via Pauli operator $\sigma_y$ of a qubit associated to the vertex $a$ of the graph $G$ of the initial graph state $\ket{G}$, yields to i) the local complementation on vertex $a$, namely, $\tau_a(\cdot)$, and the deletion of the measured vertex on the associated graph and ii) the resulting graph state $\ket{\tilde G_y} = U_{y,\pm}^{(a)}\ket{\tau_a(G)-a}$ on the unmeasured qubits. Specifically, the correction unitaries $U_{y,\pm}$ are defined according to the measurement outcome as follows:        
        \begin{align}
            \label{eq:03}
            U_{y,+} = \bigotimes_{b \in N_a}\sqrt{-i\sigma_z^{(b)}} \quad , \quad
            U_{y,-} = \bigotimes_{b \in N_a}\sqrt{i\sigma_z^{(b)}}. 
        \end{align}

    \item[III)] \textbf{Projective $\sigma_x$ Pauli measurements} \\
        The projective measurement via Pauli operator $\sigma_x$ of a qubit associated to the vertex $a$ of the graph $G$ of the initial graph state $\ket{G}$, yields to i) a sequence of local complementations on vertex $a$ and a support vertex $b_0 \in N_a$, as well as the deletion of the measured vertex on the associated graph and ii) the resulting graph state $\ket{\tilde G_x} = U_{x,\pm}^{(a)}\ket{\tau_{b_0}\left(\tau_{a}\left(\tau_{b_0}(G)\right) - a\right) }$ on the unmeasured qubits. Specifically, the correction unitaries $U_{x,\pm}$ are defined according to the measurement outcome as follows:
        \begin{align}
            \label{eq:04}
            & U_{x,+}  =
            \sqrt{i\sigma_y ^{(b_0)}}\bigotimes_{b \,\in\, N_{a}\setminus \{N_{b_0} \cup \{b_0\}\}}\sigma_z ^{(b)}, \\& 
            U_{x,-} =
            \sqrt{-i\sigma_y ^{(b_0)}}\bigotimes_{b \,\in \,N_{b_0}\setminus \{N_a \cup \{a\}\}}\sigma_z ^{(b)} \nonumber . 
        \end{align}
\end{enumerate}

The application of the Pauli projection operators reveals the output topology of the graph state after the measurement. However, in order to obtain exactly $\tilde G_i$ as output associated graph -- whose topology is described by simple graph theory tools -- it is necessary to apply the corresponding correction unitaries $U_{i,\pm}^{\dagger}$ on some unmeasured qubits, with respect to the chosen measurement basis and the measurement outcomes.
The next section provides further details.

\subsection{Introduction to QLANs}
\label{sec:2.2}

QLANs are envisioned as the building block of the future Quantum Internet \cite{MazCalCac-24, MazCalCac-24-QCNC, CheIllCac-24}. 
Unlike larger-scale quantum networks models, the implemented QLAN model relies on a centralized quantum node called \textit{orchestrator node}, which is is directly connected through a quantum channel
with each client nodes of the local area network on a physical star topology. 
The orchestrator is responsible for the generation of the initial graph state, the application of the Pauli measurements on the qubits, and the communication of the measurement outcomes to the clients. The clients on the other hand, are equipped with the minimum required quantum resources to store the entangled qubits distributed by the orchestrator and they only assume a secondary role in the network's hierarchy\footnote{Network's hierarchy refers to the allocation of significantly higher resources at the orchestrator rather than at the clients, which are lightweight.}. 
Unlike classical local area networks, the physical topology of a QLAN is not a limitation for the communication capabilities of the network, because the orchestrator can establish virtual links between clients by performing local Pauli measurements on the qubits of the generated graph state.
The reference QLAN model also assumes that after the generation of the $n$-qubit resource state, the orchestrator is able to wisely distribute a single entangled qubit for each client. At the end of the distribution process, the orchestrator retains $n_o$ qubits, and the remaining $n-n_o = k$ qubits are distributed among the clients. Therefore, the vertex set of the associated graph $G$ is partitioned into two disjoint sets, $V = V_o \cup V_c = \{o_1, \dots o_{n_o}\} \cup \{ c_1 \dots c_k \}$, where $V_o$ is the set of qubits retained by the orchestrator and $V_c$ is the set of qubits distributed among the clients. Formally:
\begin{equation}
    \label{eq:05}
    V = V_o \cup V_c, \quad V_o \cap V_c = \emptyset, \quad |V_o| = n_o, \quad |V_c| = k,
\end{equation}
thus, the total number of clients in the QLAN is exactly $k$.
As represented in Fig.~\ref{fig:02}, after the distribution process, the orchestrator node has complete control of the connectivity of the client nodes. Remarkably, by performing local Pauli measurements on its qubits, the orchestrator is able to manipulate the artificial links of the network. Hence, the chosen network logic is given by the orchestrator measurement bases, represented as a vector of the kind: $\{\sigma_{i}^{(o_1)}, \; \dots \; ,\sigma_{i}^{(o_{n_o})} \}$ with $i \in \{x,y,z\}$, that have a decisive impact on the output artificial topology.
 
\begin{figure*}[ht]
    \centering
    \includegraphics[width=0.725\linewidth]{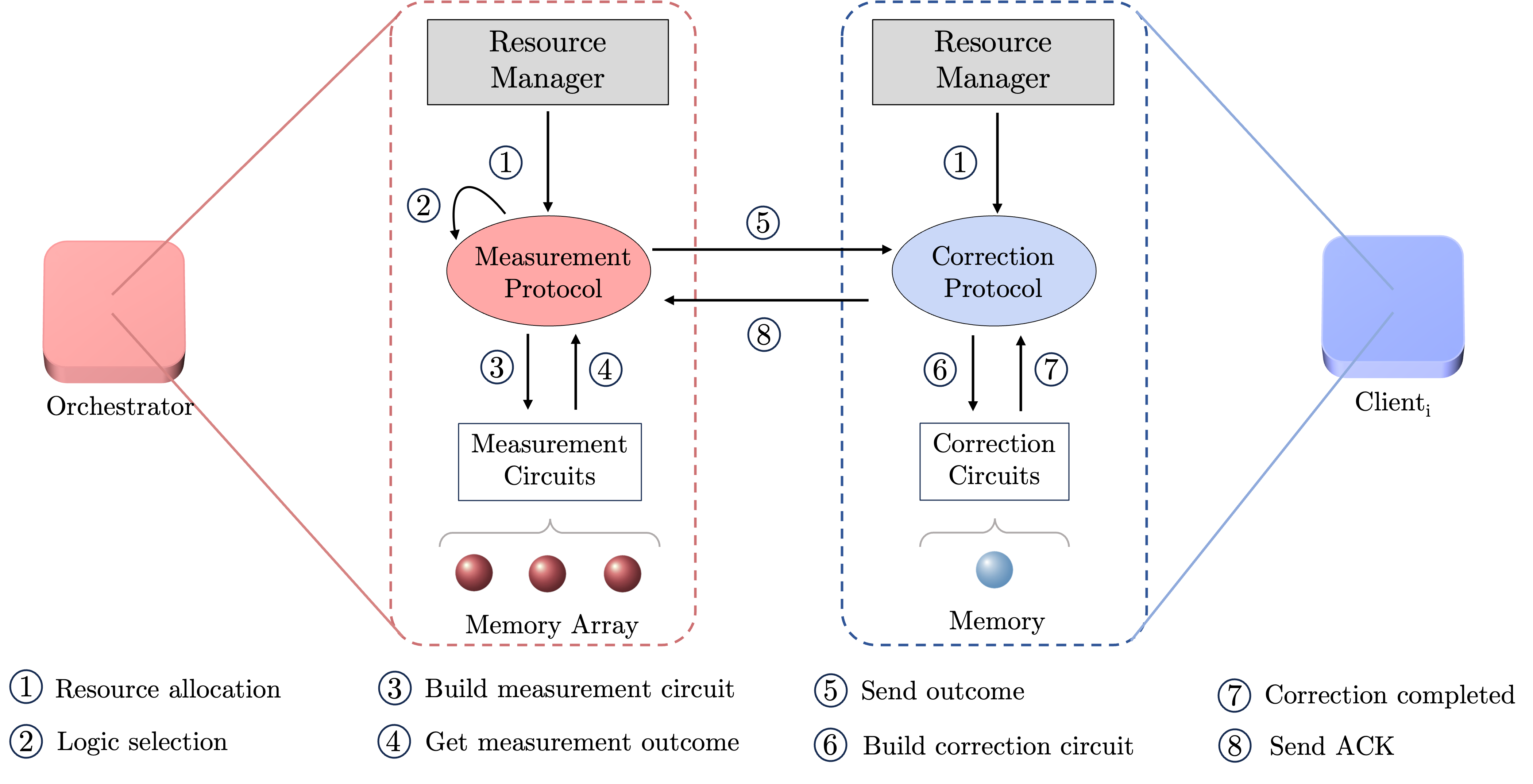}
    \caption{QLAN \textit{measurement protocol} and \textit{correction protocol} interaction. Please note that the represented interaction happens for each client of the network. A detailed description of the implementation of each protocol is given in Alg.\ref{alg:01} and \ref{alg:02}.} 
    \label{fig:03}
    \hrulefill
\end{figure*}

\section{QLAN Module in SeQUeNCe}
\label{sec:3}

\begin{algorithm}[t!]
    \caption{Measurement Protocol}
    \label{alg:01}
    \For{$\sigma_i^{(j)} \in$ \textit{bases-array} }{
        \textcolor{Orchid}{$\mathcal{Apply}$ $P_i$ $\mathcal{on \; qubit}$ $o_j$} \\
        {\textit{meas-outcomes(j)} $\gets m_j$}
    }
    \For{$m_j \in \textit{meas-outcomes}$}{
    
    \uIf{\textit{bases-array(j)} is $\sigma_x $}{ 
        {$b_0 \gets N_{o_{j}}[0]$} \\
        {$N_{b_0} \gets \{\bar o \in V_o: \exists (b_0, \bar o) \in E_{\bar o}\}$} \\
        {\textit{MsgType} $\gets$ \textit{B0\_Designation}} \\
        \textcolor{PineGreen}{\texttt{Send \textit{MsgType} to $b_0$ node}} \\

        \uIf{$m_j$ is "$+$"}
        {
        {\textit{dest-sample} $\gets \{b_0\} \cup \{N_{o_j} \setminus (N_{b_0} \cup {b_0})\}$} \\ 
        {\textit{MsgType} $\gets \ket{x,+}$} 
        }
        \uElseIf{$m_j$ is "$-$"}{
            {\textit{dest-sample} $\gets \{b_0\} \cup \{N_{b_0} \setminus (N_{o_j} \cup {o_j})\}$} \\ 
            {\textit{MsgType} $\gets \ket{x,-}$} \\
            
        }
        \textcolor{PineGreen}{\texttt{Send \textit{MsgType} to \textit{dest-sample}}} \\
        \uIf{$b_0 \in N_{o_{j+1}}$}{
            $N_{o_{j+1}} \gets  (N_{o_j} \cup N_{o_{j+1}}) \setminus (N_{o_j} \cap N_{o_{j+1}})$
        }
    }
    \uElse{ 
        {\textit{dest-sample} $\gets N_{o_{j}}$} \\
        \uIf {\textit{bases-array(j)} is $\sigma_y$}{
            \uIf {$m_j$ is "$+$"}{
            {\textit{MsgType} $\gets \ket{y,+}$}
            }
            \uElseIf {$m_j$ is "$-$"}{
            {\textit{MsgType} $\gets \ket{y,-}$} 
            }
        \textcolor{PineGreen}{\texttt{Send \textit{MsgType} to \textit{dest-sample}}}
        }
        
        \uElseIf{\textit{bases-array(j)} is $\sigma_z$}{
            \uIf {$m_j$ is "$+$"}{
                {\textit{MsgType} $\gets \ket{z,+}$}
            }
            \uElseIf {$m_j$ is "$-$"}{
                {\textit{MsgType} $\gets \ket{z,-}$}
            }
        \textcolor{PineGreen}{\texttt{Send \textit{MsgType} to \textit{dest-sample}}}
        }
    }
    {$N_{o_{j}} \gets \emptyset $}
    }
\end{algorithm}

\begin{algorithm}[t!]
    \caption{Correction Protocol}
    \label{alg:02}
        \While{ \textcolor{PineGreen}{\texttt{new message arrives}}}{

            \Switch{\textit{MsgType}}{
            
            \uCase{\textit{MsgType} is 
                \textit{B0\_Designation}}{
                $b_0 \gets \text{True}$
            }

            \uCase{\textit{MsgType} is $\ket{z,+}$}{
                \textcolor{Orchid}{$\mathcal{No \; correction  \; is \; needed}$}
            }

            \uCase{\textit{MsgType} is $\ket{z,-}$}{
                \textcolor{Orchid}{$\mathcal{Apply}$ $\sigma_z$ $\mathcal{gate \; on \; its \; own \; qubit}$}
            }
            \uCase{\textit{MsgType} is $\ket{y,+}$}{ \textcolor{Orchid}{$\mathcal{Apply}$ $\sqrt{-i\sigma_z} ^{\dagger}$ $\mathcal{on \;its  \; own \; qubit}$}
            }

            \uCase{\textit{MsgType} is $\ket{y,-}$}{
            \textcolor{Orchid}{$\mathcal{Apply}$ $\sqrt{i\sigma_z} ^{\dagger}$ $\mathcal{on \;its \; own \;qubit}$}
            }

            \uCase{\textit{MsgType} is $\ket{x,+}$}{
            \uIf{$b_0 \text{ is True}$}{
                    \textcolor{Orchid}{$\mathcal{Apply}$ $\sqrt{i\sigma_y} ^{\dagger}$ $\mathcal{on \; its \; own \; qubit}$}
                    $b_0 \gets \text{False}$
                }
            \uElse{
                    \textcolor{Orchid}{$\mathcal{Apply}$ $\sigma_z$ $\mathcal{gate \;on \;its \;own \;qubit}$}
                }
            }

            \uCase{\textit{MsgType} is $\ket{x,-}$}{
            \uIf{$b_0$ is \text{True}}{
                    \textcolor{Orchid}{$\mathcal{Apply}$ $\sqrt{-i\sigma_y} ^{\dagger}$ $\mathcal{on \;its \;own \;qubit}$}
                    $b_0 \gets \text{False}$
                }
            \uElse{
                    \textcolor{Orchid}{$\mathcal{Apply}$ $\sigma_z$ $\mathcal{gate \;on \;its \;own \;qubit}$}
                }
            }
            \textcolor{PineGreen}{\texttt{Send ACK message to Orchestrator}}
            }
        }
        \vspace{-0.75ex}
\end{algorithm}

A complete simulation of the QLAN model proposed in \cite{MazCalCac-24-QCNC,MazCalCac-24} can be implemented in SeQUeNCe according to the main modules of the simulator. Specifically, the network nodes are implemented as \textit{QuantumNode} entities, which are able to store qubits and perform quantum operations. We customized SeQUeNCe's QuantumNode into two new types: the \textit{OrchestratorNode} and the \textit{ClientNode} that represent the different physical requirements (i.e., the node hierarchy) for the network nodes. 

\subsection{QLAN Module Design}

\begin{itemize}
    \item \textbf{Orchestrator node}: the orchestrator node requires an array of quantum memories to store the qubits of the generated graph state. Moreover, it is equipped with a set of quantum gates to perform the desired measurement circuits. Specifically, the execution of the measurement circuits requires the orchestrator to coordinate its operations -- according to the desired network logic -- with a \textit{measurement protocol}. Finally, it also requires a \textit{Resource Manager}\footnote{Each network node implemented in SeQUeNCe is provided with a resource manager to allocate and manage the resources needed for the protocols.}, that is responsible for the management and generation of the processes related to this node.
    \item \textbf{Client node}: each client node requires -- for the sake of simplicity --  a single quantum memory to store the qubits distributed by the orchestrator node. Moreover, it is equipped with a set of quantum gates able to implement correction circuits. A client node requires a \textit{correction protocol} to manage the correct application of the corrections stemming from the outcomes received from the orchestrator. Finally, similarly to the orchestrator node, it also requires a Resource Manager.
\end{itemize}

As discussed in Sec.~\ref{sec:2.2}, for the sake of simulating the functionalities provided by the QLAN model, we require a $n = n_o + k$ qubit graph state to be distributed among the network nodes. As instance, the simplest state\footnote{We emphasize that every graph state satisfying \eqref{eq:05} can be tested with the module.} to be generated and distributed is a linear graph state, namely, a graph state where each qubit is connected to the next one and $n_o = k-1$.
For the sake of the simulation, the circuits required for the generation of the required simulation can be fully simulated in SeQUeNCe, with a given number of clients. Specifically, the simulation of quantum gates in SeQUeNCe is implemented using \textit{QuTiP} \cite{qutip2}, an open source simulation tool widely adopted by the quantum simulation community.

The simulation of the QLAN model is based on the execution of two core protocols, the \textit{measurement protocol} and the \textit{correction protocol} running at the orchestrator node and the client nodes, respectively. As depicted in Fig.~\ref{fig:03}, the implemented protocols rely on adequately performing quantum operations -- represented by quantum circuits -- and classical signaling in order to correctly apply the desired network logic. More in detail, the classical signaling between the two protocols uses a classical channel that connects each client node to the orchestrator in a star topology. Moreover, in order to emphasize the interplay between classical and quantum operations performed at the network nodes, the protocols are described in Alg.~\ref{alg:01} and \ref{alg:02}, respectively\footnote{The definition of protocol adopted here does not take into account the formal definition of i) the syntax of the messages and ii) semantics fields. Indeed, the classical message is modeled through the \textit{MsgType} variable in the pseudo-code.}. We use different colors and fonts for denoting \textcolor{Orchid}{\textsc{$\mathcal{local\;quantum\;operations}$}} and \textcolor{PineGreen}{\texttt{classical communication}} (LOCC) involved in the protocols. For the sake of brevity, we assume that the orchestration logic (i.e., the bases array) has already been decided by the orchestrator.

\subsection{Measurement Protocol}
\label{sec:3.1}
The measurement protocol represents the sequence of rules and messages exchanged by the Orchestrator Node to perform the desired topology engineering process. 
Specifically, by appropriately measuring its own qubits, the orchestrator updates accordingly the set of destinations for the measurement outcomes. More in detail, the orchestrator's operations are described by the \textit{bases-array}, which determines both the measurement circuits to be performed and the classical messages to be sent to the clients. As detailed in Alg.~\ref{alg:01}, the set of destination clients -- referred to as \textit{dest-sample} -- for the measurement outcomes strictly depends on the measurement basis and the measurement outcomes -- according to Eq.~\eqref{eq:02},~\eqref{eq:03} and~\eqref{eq:04} --  but are also related to the neighborhood of the $j$-th qubit measured at the orchestrator, namely, $N_{o_j}$. This consideration leads to the potential arrival of multiple correction operations at the same client belonging to different orchestrator's qubit neighborhood $N_{o_j}$. Moreover, according to the arbitrary choice of the $b_0$ node, a $\sigma_x$ measurement can cause an update of the neighbors of the following orchestration qubit. This specific behavior, is referred to as rolling effect \cite{MazCalCac-24}.
Moreover, as shown in the last line of Alg.~\ref{alg:01}, at each measurement operation with any chosen base, the neighborhood of the measured qubit is also left empty. 

\begin{figure*}[t!]
    \centering
    \includegraphics[width=0.80\linewidth]{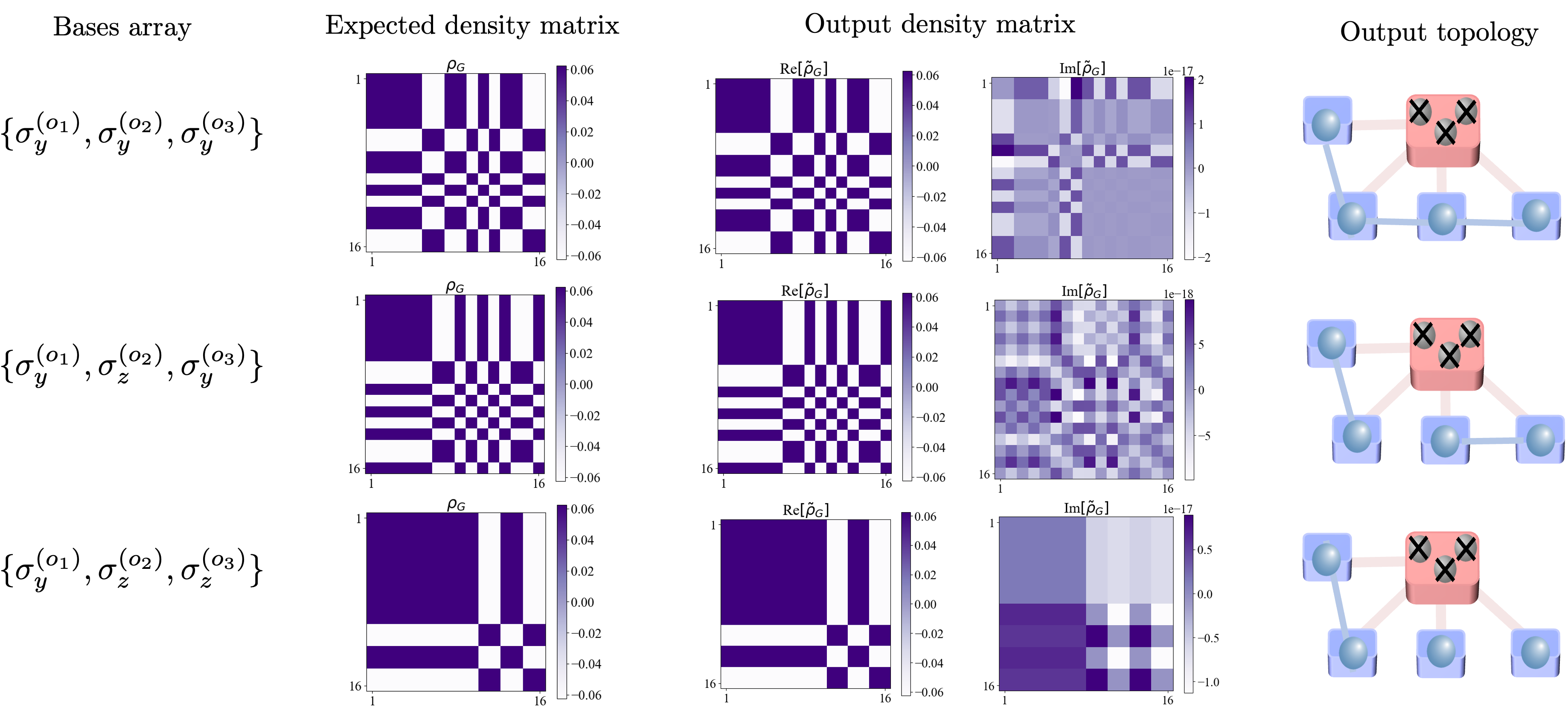}
    \caption{Resulting graph states in a $4$-clients topology after the application of Pauli measurements and corrections.}
    \label{fig:04}
    \hrulefill
\end{figure*}

\subsection{Correction Protocol}
\label{sec:3.2}
As described in Alg.~\ref{alg:02}, the correction protocol starts when a client receives a classical message from the orchestrator. Each message represents the type of measurement that was performed and its outcome. With this information, the client is then responsible for applying the corresponding correction unitaries, according to Eqs.~\eqref{eq:02},~\eqref{eq:03} and \eqref{eq:04}.

\begin{remark}
    The corrections depend on the measurement outcomes and the measurement bases, as detailed in Sec.~\ref{sec:2.2}. Remarkably, it is sufficient for the clients to apply the gate corresponding to the complex conjugate $U_{i,\pm} ^{\dagger}$  to cancel the effects of the $U_{i,\pm}$ unitaries.
\end{remark}

Recall that for the correction process to be successful, all clients must be listening for potential messages from the orchestrator, therefore it is very likely that several instances of the correction protocol will run in parallel on several clients.

\subsection{Module Validation}
The module has been validated with a physical star topology equipped with a single orchestrator and an arbitrary number of clients. 
Specifically, the considered simulations start when the client and orchestrator memories correctly store the qubits of graph states corresponding to the initial topology. We also note that the module's logic -- with a different resource state generation -- also holds for an arbitrary graph state generated at the orchestrator and satisfying the conditions presented in \eqref{eq:05}. As a result, by comparing the resulting density matrices $\tilde \rho_G = \ket{\tilde G}\bra{\tilde G}$ with the expected ones $\rho_G$, we note that the resulting states, obtained according to the desired orchestration logic, perfectly match with the theoretical prediction -- by assuming that the undesired imaginary part is completely negligible -- in the absence of noise. Fig.~\ref{fig:04} presents the simulation results of a 4-clients QLAN topology\footnote{The mean simulation time for a 4-client topology on a M2 Macbook Pro is $0.028$s with a standard deviation of $0.011$s. Simulation time grows in the order of seconds for simulations with at least 6 clients.}, where the final graph state obtained after of the corrections coincides with the desired artificial topology. Remarkably, the desired orchestration logic can be applied with an arbitrary choice of measurement bases and with all the possible outcomes.  

\section{Conclusion}
\label{sec:conclusion}
In this paper, we discussed the design and implementation of the functionalities of the QLAN model proposed in \cite{MazCalCac-24,MazCalCac-24-QCNC} using SeQUeNCe. Specifically, we described the design of the implemented software module and the functionalities that can be simulated, with a focus on the coordination and the interplay between quantum operations and classical signaling. More in detail, in order to reproduce the centralized model of QLAN, a \textit{measurement protocol} is required at the orchestrator and a \textit{correction protocol} is required at each client in the network. Thanks to the implementation of the network nodes' unique features and protocols, it has been possible to validate the theoretical framework of the QLAN model by comparing the theoretical results with the output topologies. The simulation matches the theoretical prediction with an arbitrary Pauli measurement array and with a given number of clients.

\bibliography{biblio}
\bibliographystyle{ieeetr}

\end{document}